\documentclass[referee, sn-standardnature]{sn-jnl}
\usepackage[utf8]{inputenc}
\usepackage{graphicx}	% Including figure files
\usepackage{amsmath}	% Advanced maths commands
\usepackage{amssymb}	% Extra maths symbols
\usepackage{xcolor}
\usepackage{amsthm}    
\def \aap {A\&A} % alternative A&A code

\def \aj {AJ}

\def \apj {ApJ}
\def \apjl {ApJL}
\def \apjs {ApJS}
\def \aaps {Ap\&SS}
\def \mnras {MNRAS}
\def \nat {Nat}
\def \pasp {PASP}

\def \ssr {Space Sci.\ Rev.}

\title[Polarized X-rays from Mrk 501]{Polarized Blazar X-rays imply particle acceleration in shocks}

\author*[1]{Ioannis Liodakis}
\affil*{yannis.liodakis@utu.fi}
\author[2]{Alan P. Marscher}
\author[3]{Iv\'{a}n Agudo}
\author[4]{Andrei V. Berdyugin}
\author[3]{Maria I. Bernardos}
\author[3,5]{Giacomo Bonnoli}
\author[6]{George A. Borman}
\author[7,8]{Carolina Casadio}
\author[3]{V\'{i}ctor Casanova}
\author[9]{Elisabetta Cavazzuti}
% ...

\affil[1]{Finnish Centre for Astronomy with ESO, FI-20014 University of Turku, Finland}
\affil[2]{Institute for Astrophysical Research, Boston University, 725 Commonwealth Avenue, Boston, MA 02215, USA}
\affil[3]{Instituto de Astrof\'{i}sica de Andaluc\'{i}a, IAA-CSIC, Glorieta de la Astronom\'{i}a s/n, 18008 Granada, Spain}
\affil[4]{Department of Physics and Astronomy, FI-20014 University of Turku, Finland}
\affil[5]{INAF Osservatorio Astronomico di Brera, Via E. Bianchi 46, 23807 Merate (LC), Italy}
\affil[6]{Crimean Astrophysical Observatory RAS, P/O Nauchny, 298409, Crimea}
\affil[7]{Institute of Astrophysics, Foundation for Research and Technology - Hellas, Voutes, 7110 Heraklion, Greece}
\affil[8]{Department of Physics, University of Crete, 70013, Heraklion, Greece}
\affil[9]{Agenzia Spaziale Italiana, Via del Politecnico snc, 00133 Roma, Italy}
% ...

\begin{document}

\maketitle

{\bf 
Most of the light from blazars, active galactic nuclei with jets of magnetized plasma that point nearly along the line of sight, is produced by high-energy particles, up to $\sim 1$ TeV. Although the jets are known to be ultimately powered by a supermassive black hole, how the particles are accelerated to such high energies has been an unanswered question. The process must be related to the magnetic field, which can be probed by observations of the polarization of light from the jets. Measurements of the radio to optical polarization - the only range available until now - probe extended regions of the jet containing particles that left the acceleration site days to years earlier \cite[e.g.,][]{Jorstad2005,Marin2018,Blinov2021}, and hence do not directly explore the acceleration mechanism, as could X-ray measurements. Here we report the detection of X-ray polarization from the blazar Markarian~501 (Mrk~501). We measure an X-ray linear polarization degree $\Pi_X \sim10\%$, a factor of $\sim2$ higher than the value at optical wavelengths, with a polarization angle parallel to the radio jet. This points to a shock front as the source of particle acceleration, and also implies that the plasma becomes increasingly turbulent with distance from the shock.}

In blazars whose lower-energy emission component peaks in the X-ray band like Mrk~501, synchrotron radiation is the dominant emission process from radio to X-rays.  Radiation at longer wavelengths likely arises from larger regions in the jet, hence multiwavelength studies probe spatial variations in the magnetic field structure and other physical properties in different locations \cite{Marscher1985,Marscher2014}. A particularly important diagnostic is the degree of order of the magnetic field and its mean direction relative to the jet axis, which can be determined by measurements of the linear polarization. For example, particle acceleration at a shock front should result in relatively high levels (tens of percent) of X-ray linear polarization along a position angle that is parallel to the jet \cite{Tavecchio2018}. In contrast, more stochastic acceleration processes involving turbulence or plasma instabilities are expected to lead to weak polarization with random position angles. The optical, infrared, and radio polarization probe the level of order and mean direction of the magnetic field in regions progressively farther from the site of particle acceleration. Simultaneous multiwavelength polarization from X-ray to radio, now possible with the advent of the Imaging X-ray Polarimetry Explorer ({\it IXPE}, \cite{Weisskopf2022}), can therefore provide a more complete picture of the emission region of a blazar jet than has been possible until now.

Variations in the flux of blazars at all wavebands, and in the linear polarization at radio to optical wavelengths, is largely stochastic in nature, which can be interpreted as the result of turbulence \cite{Marscher2014,Peirson2018,Tavecchio2018,Peirson2019}. Multi-zone emission models, often involving a turbulent magnetic field, can reproduce a number of the observed characteristics of the variable linear polarization. In a turbulent region, roughly modeled as $N$ cells, each with a uniform but randomly oriented field, we expect we expect a mean degree of polarization $\langle \Pi \rangle \sim 75\%/\sqrt{N}$, with the value of $\Pi$ exhibiting variability on short time-scales with a standard deviation  $\sim 0.5\langle\Pi\rangle$ \cite{Marscher2014}, as often observed \cite{MJ2021}. For a turbulent field in the plasma crossing a shock front, particle acceleration should be most efficient in cells whose magnetic field is nearly parallel to the shock normal; this bias leads to a higher value of $\Pi$ and more pronounced variability at X-rays compared to lower frequencies \cite{Marscher2014}. The passage of turbulent cells through the emission region would also cause irregular variations, including some apparent rotations, in $\psi$ \cite{Marscher2008,Peirson2018}.

On the other hand, some of the observed radio and optical patterns of polarization variability (e.g., the aforementioned $\psi$ rotations) have been found to be inconsistent with purely stochastic processes \cite{Kiehlmann2017,Blinov2018}. This indicates that there is some coherent ordering of the magnetic field, e.g., by compression or amplification by plasma processes in shocks \cite[e.g.,][]{Tavecchio2020} or by the presence of a global, perhaps helical, magnetic field component \cite[e.g.,][]{Lyutikov2005,Hovatta2012,Gabuzda2021}. In the commonly used single zone model, the radiating particles are accelerated by an unspecified process to highly relativistic energies while confined within a plasmoid with a partially ordered or helical magnetic field. The global magnetic field structure is expected to produce similar polarization patterns across frequencies, with little variability over time \cite{DiGesu2022}. If the field is helical, $\psi$ should align with the jet direction for most viewing angles \cite{Lyutikov2005}. In an alternative scenario, which includes shock acceleration, particles become energized over a limited volume - e.g., at a shock front - and then advect or diffuse away from that region \cite{Marscher1985,Angelakis2016,Tavecchio2018}. In this process, the electrons lose energy to radiation, and so emit at progressively decreasing frequencies as they travel away from the acceleration site. We refer to this model as ``energy-stratified''. If the magnetic field is well ordered over the small volume of the acceleration region and becomes increasingly turbulent farther downstream, $\Pi$ will decrease toward longer wavelengths, while $\psi$ can vary with frequency if the mean direction of the magnetic field changes as the volume increases. In Mrk~501, we expect a progressively higher $\Pi$ from radio to X-rays. A shock partially orders the magnetic field of the plasma crossing the shock, with the ordered field perpendicular to the shock normal. This causes the net polarization electric vector to be aligned with the jet. In a kink-instability induced magnetic reconnection scenario, where contiguous regions of oppositely-directed magnetic field come into contact,  the jet flow is sheared because of transverse velocity gradients \cite{Sironi2021}. Shearing would stretch the magnetic field along the jet boundary, so that $\psi$ is expected to be transverse to the jet direction. The simultaneous contribution of multiple current sheets will lead to overall lower polarization than in a shock scenario, with similar levels of polarization across frequencies \cite[e.g.,][]{Tavecchio2021}. Our expectations from the different emission models are summarized in Table  \ref{tab:scenarios}.

\begin{figure}
\centering
\includegraphics[scale=0.3]{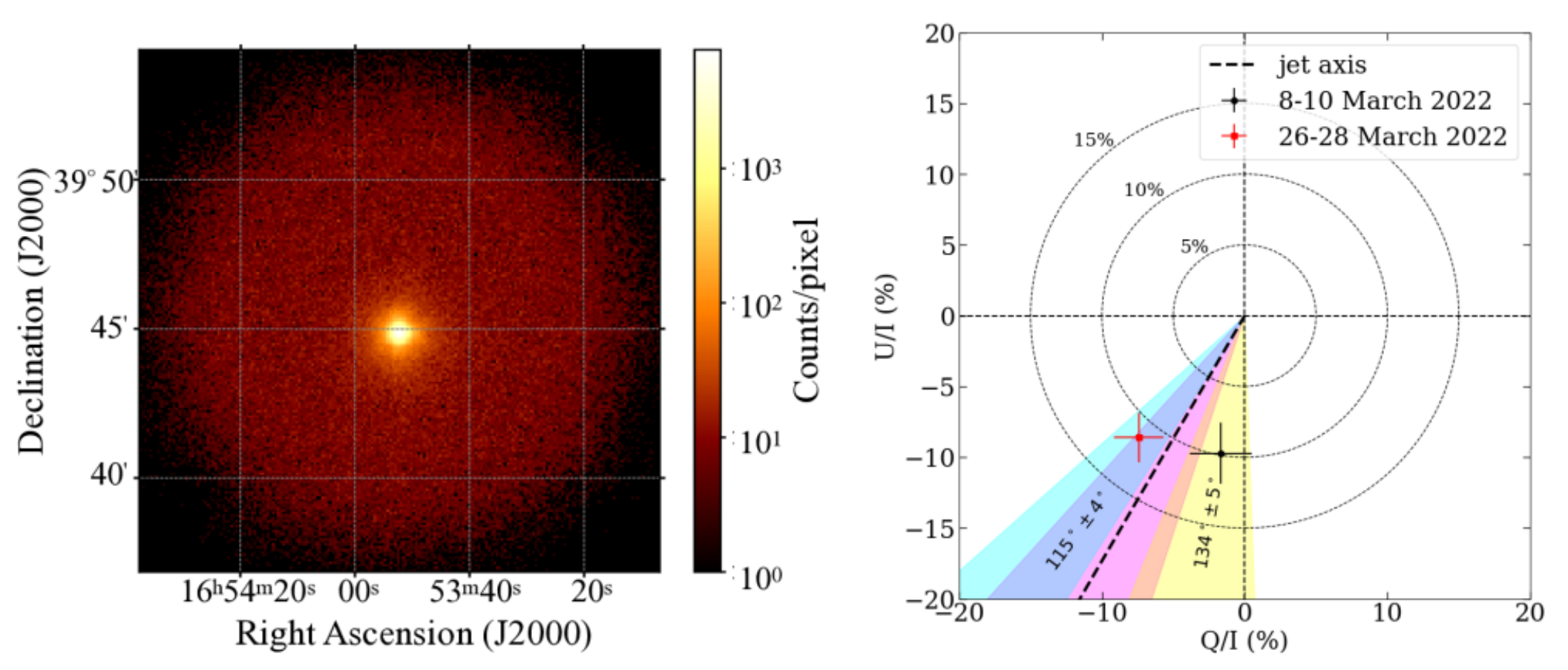}
 \caption{{\it IXPE} observations of Mrk~501. {\bf Top left:} {\it IXPE} image of Mrk 501 during the 8-10 March 2022 observation in the 2-8~keV band. The colorbar denotes the number of X-ray photons per pixel. 
 {\bf Top right:} Normalized Stokes Q and Stokes U parameters of both {\it IXPE} observations. The yellow and cyan shaded regions denote the uncertainty (68\% CI) in the polarization angle for the 8-10 March and 26-28 March observations respectively. The dashed black line shows the jet direction and the magenta shaded area its uncertainty (68\% CI). The dashed circles mark different levels of polarization degree, as labeled. Error bars denote the 68\% CI.}
\label{plt:ixpe_data}
\end{figure}

\begin{figure}
\centering
 \includegraphics[scale=0.2]{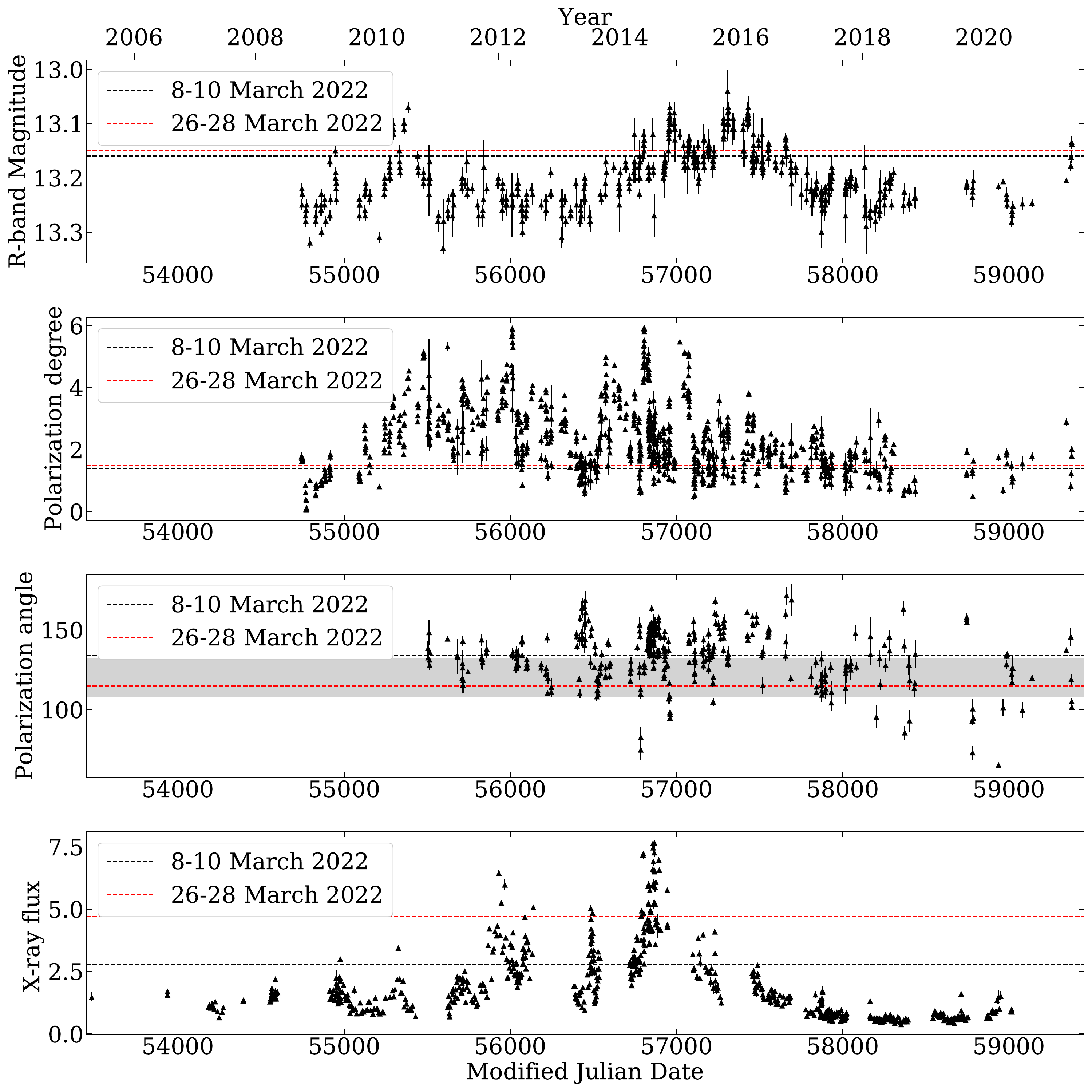}
 \caption{Multiwavelength and polarization archival observations of Mrk~501. Optical brightness (R-band, upper panel), observed optical $\Pi$ in \% (second-from-top panel), observed optical $\psi$ in degrees (third-from-top panel), and X-ray flux in $\times10^{-10}\rm erg/s/cm^2$ (lower panel). The black and red dashed line indicate the level of the source during the 8-10 March and 26-28 March {\it IXPE} observations respectively. The grey shaded area in the $\psi$ shows the direction of the jet axis. In all panels the error bars denote the 68\% CI.}
\label{plt:light_mrk501}
\end{figure}

%Here, we report the detection of X-ray polarization from the blazar Markarian 501 (Mrk 501) by {\it IXPE}.
The first {\it IXPE} observation of Mrk~501 took place during 2022 March 8-10 (100~ksec, MJD 59646-59648) and was accompanied by observations across the electromagnetic spectrum from multiple observatories (see Methods). {\it IXPE} measured a polarization degree $\rm\Pi_X=10\pm2\%$ and an electric vector position angle $\rm\psi_X=134^\circ\pm5^\circ$ (measured East of North) over the X-ray energy range of 2-8 keV. Contemporaneous radio-millimeter and optical observations (Extended Data Table \ref{tab:mult_obs}) measured $\rm\Pi_R=1.5\pm0.5\%$ along $\rm\psi_R=152^\circ\pm10^\circ$ and  $\rm\Pi_O=4\pm1\%$ along  $\rm\psi_O=119^\circ\pm9^\circ$. A second {\it IXPE} observation took place 2022 March 26-28 (86~ksec, MJD 59664-59667) yielding  $\rm\Pi_X=11\pm2\%$ along $\rm\psi_X=115^\circ\pm4^\circ$. Simultaneously to the second observation, the optical polarization was measured as $\rm\Pi_O=5\pm1\%$ along $\rm\psi_O=117^\circ\pm3^\circ$ (Extended Data Table \ref{tab:mult_obs2}). The two observed $\rm\psi_X$ are consistent within 3$\sigma$. The radio and optical $\rm\psi$ also lie within 3$\sigma$ from each other and $\rm\psi_X$. Moreover, the position angle of Mrk~501's jet has been determined through Very Long Baseline Array (VLBA) imaging at 43~GHz to be $120^\circ\pm12^\circ$ \cite{Weaver2022}. This would suggest that, in both cases, radio-to-X-ray $\psi$ is aligned with the jet axis within uncertainties (Fig. \ref{plt:ixpe_data}). We do not find evidence of polarization variability during either {\it IXPE} observation. Compared to the archival multiwavelength observations, we find the flux and polarization of Mrk 501 for both observations to be within one standard deviation of the median of the long-term light curves (Fig. \ref{plt:light_mrk501}). Blazars like Mrk~501 are known to reach as much as an order of magnitude higher X-ray fluxes during outbursts. For the first {\it IXPE} observation the measured X-ray flux indicates an average activity state, while during the second observation we find evidence of a slightly elevated X-ray flux state. Compared to the historical maximum X-ray flux, during our observations Mrk~501 was a factor of three and a factor of two fainter, respectively.

\begin{table}
\centering
\caption{Summary of model properties. We find increasing $\Pi$ towards higher frequencies, no significant variability during the 2-3 day long {\it IXPE} observations, and rough alignment of $\psi$ with the jet axis from radio to X-rays. Therefore, a shock-accelerated, energy-stratified electron population model satisfies all our multiwavelength polarization observations.}
\begin{tabular}{cccc}
\hline
Model & Multiwavelength & X-ray polarization & X-ray polarization\\
 & polarization & variability$^\dagger$ &  angle\\
\hline
Single-zone & constant$^*$ & slow & any \\
%\hline
Multi-zone & mildly chromatic & high & any \\
%\hline
 Energy stratified &  strongly chromatic & slow & along the \\
 (shock) & &  & jet axis \\
%\hline
Magnetic reconnection & constant & moderate & perpendicular \\
(kink instability) &  &  &  to jet axis \\
\hline
Observed &  strongly chromatic & slow &  along the\\
 & &  & jet axis \\
 \hline
\end{tabular}
\tablenotes{$^*$There is a slight dependence on the slope of the emission spectrum.}
\tablenotes{$^\dagger$Slow variability = a few days to week, moderate variability = days, high variability $\leq 1$ day.}
\label{tab:scenarios}
\end{table}

\begin{figure}
\centering
\includegraphics[scale=0.23]{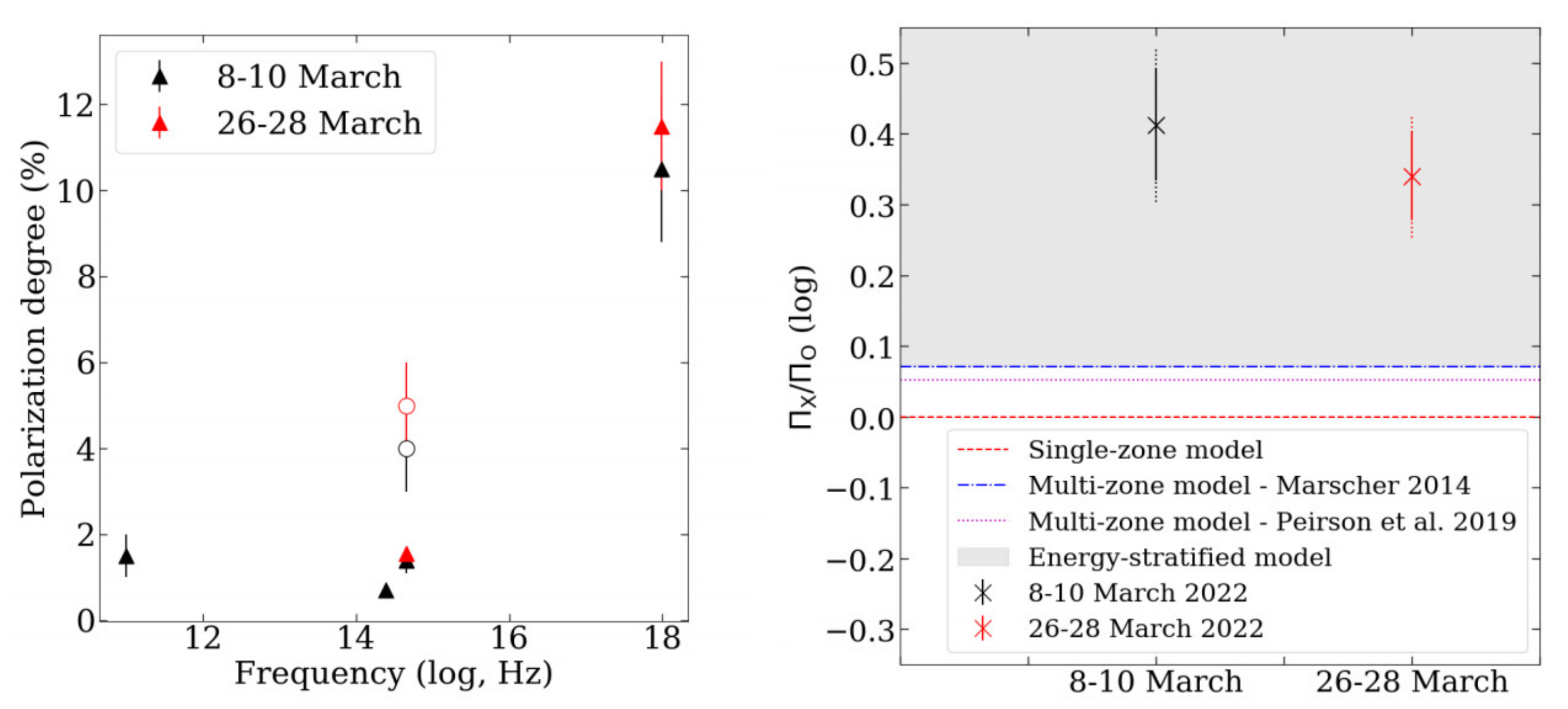}
 \caption{Multiwavelength polarization of Mrk~501. {\bf Left:} Multiwavelength polarization degree of Mrk~501 from radio to X-rays. Black symbols are for the 8-10 March observation, and red for the 26-28 March observation. The open symbols show the host-galaxy corrected -- intrinsic optical polarization degree. {\bf Right:} Comparison between the observed logarithm of the X-ray and optical $\Pi$ ratio and the expectations from single-zone (red dashed line), two turbulent multi-zone jet models (dash-dotted blue and dotted magenta lines), and energy-stratified models (grey shaded area) for both {\it IXPE} observations (black for 8-10 March and red for 26-28 March). The solid errorbars show the ratio uncertainty from the {\it IXPE} measurements; the dotted errorbars show the full uncertainty including optical uncertainties. In both panels the error bars denote the 68\% CI.}
\label{plt:polssed_mrk501}
\end{figure}

The polarization measurements reported here reveal an increase in $\Pi$ toward higher frequencies, with a degree of X-ray polarization that is more than twice the optical value (Fig. \ref{plt:polssed_mrk501}). This is in tension with the single-zone, turbulent multi-zone, and magnetic reconnection models discussed above. There is no significant variability within the duration of the individual {\it IXPE} observations, contrary to the predicted behaviour if turbulent cells moved in and out of the emission region on time-scales of $\lesssim 2$ days. On the other hand, the low ($<10\%$) optical and X-ray polarization suggests significant disordering of the local magnetic field, possibly due to the presence of stationary turbulence. The wavelength dependence and lack of variability of $\Pi$, plus constancy of $\psi$ and its alignment with the jet direction, supports the shock-accelerated energy-stratified electron population scenario \cite{Marscher1985,Angelakis2016,Tavecchio2021}.  Previous intensely-sampled measurements of the polarization of Mrk~501 have found variations in $\rm\Pi_O$ by $\pm5\%$ and in $\rm\psi_O$ by $\sim50^\circ$ from one night to the next \cite{MJ2021}. These apparently discrepant results can be reconciled if the turbulence of the plasma flowing through shocks in the jet is only intermittent, as has been found previously in other blazars \cite{Webb2021}. One would also expect deviations of the observed $\rm\psi$ from the jet axis as one moves further away from the shock front into more turbulent regions of the jet. At present, the large $\rm\psi$ uncertainties prevent us from confirming such behavior. Future observations of Mrk~501 or similar blazars will allow us to explore the jet's multiwavelength polarization variability. A prediction of the energy-stratified model is that the X-ray polarization angle of blazars whose synchrotron spectral energy distribution peaks at X-ray frequencies, like Mrk~501, will exhibit rotations \cite{Blinov2016}.

Probing the magnetic environment of the radiating particles site of energization has supplied a new method for discriminating among particle acceleration mechanisms in astrophysical jets. The new X-ray polarization observations, in combination with the previously available radio and optical polarization diagnostics, have provided a discriminating set of evidence. Our results demonstrate how multiwavelength polarization uniquely probes the physical conditions in supermassive black-hole systems. Future monitoring of the time variability of multiwavelength polarization with {\it IXPE} and other instruments will define better the range of physical conditions that occur in astrophysical jets.

\bmhead{Acknowledgments}

I.L. thanks the Kavli Institute for the Physics and Mathematics of the Universe for their hospitality while this paper was written. The authors thank A. Veledina for discussions that helped improve this work. I.L. was supported by the JSPS postdoctoral short-term fellowship program. The Imaging X ray Polarimetry Explorer (IXPE) is a joint US and Italian mission.  The US contribution is supported by the National Aeronautics and Space Administration (NASA) and led and managed by its Marshall Space Flight Center (MSFC), with industry partner Ball Aerospace (contract NNM15AA18C).    The Italian contribution is supported by the Italian Space Agency (Agenzia Spaziale Italiana, ASI) through contract ASI-OHBI-2017-12-I.0, agreements ASI-INAF-2017-12-H0 and ASI-INFN-2017.13-H0, and its Space Science Data Center (SSDC) with agreements ASI-INAF-2022-14-HH.0 and ASI-INFN 2021-43-HH.0, and by the Istituto Nazionale di Astrofisica (INAF) and the Istituto Nazionale di Fisica Nucleare (INFN) in Italy.  This research used data products provided by the IXPE Team (MSFC, SSDC, INAF, and INFN) and distributed with additional software tools by the High-Energy Astrophysics Science Archive Research Center (HEASARC), at NASA Goddard Space Flight Center (GSFC). Data from the Steward Observatory spectropolarimetric monitoring project were used. This program is supported by Fermi Guest Investigator grants NNX08AW56G, NNX09AU10G, NNX12AO93G, and NNX15AU81G. This research has made use of data from the RoboPol programme, a collaboration between Caltech, the University of Crete, IA-FORTH, IUCAA, the MPIfR, and the Nicolaus Copernicus University, which was conducted at Skinakas Observatory in Crete, Greece. The IAA-CSIC co-authors acknowledge financial support from the Spanish "Ministerio de Ciencia e Innovacion (MCINN) through the "Center of Excellence Severo Ochoa" award for the Instituto de Astrof\'{i}sica de Andaluc\'{i}a-CSIC (SEV-2017-0709). Acquisition and reduction of the POLAMI and MAPCAT data was supported in part by MICINN through grants AYA2016-80889-P and PID2019-107847RB-C44. The POLAMI observations were carried out at the IRAM 30m Telescope. IRAM is supported by INSU/CNRS (France), MPG (Germany) and IGN (Spain). The research at Boston University was supported in part by National Science Foundation grant AST-2108622, NASA Fermi Guest Investigator grant 80NSSC21K1917, and NASA Swift Guest Investigator grant 80NSSC22K0537. This study uses observations conducted with the 1.8 m Perkins Telescope Observatory (PTO) in Arizona (USA), which is owned and operated by Boston University. Based on observations obtained at the Hale Telescope, Palomar Observatory as part of a continuing collaboration between the California Institute of Technology, NASA/JPL, Yale University, and the National Astronomical Observatories of China. This research made use of Photutils, an Astropy package for detection and photometry of astronomical sources (Bradley et al., 2019). GVP acknowledges support by NASA through the NASA Hubble Fellowship grant  \#HST-HF2-51444.001-A  awarded  by  the  Space Telescope Science  Institute,  which  is  operated  by  the Association of Universities for Research in Astronomy, Incorporated, under NASA contract NAS5-26555. Based on observations made with the Nordic Optical Telescope, owned in collaboration by the University of Turku and Aarhus University, and operated jointly by Aarhus University, the University of Turku and the University of Oslo, representing Denmark, Finland and Norway, the University of Iceland and Stockholm University at the Observatorio del Roque de los Muchachos, La Palma, Spain, of the Instituto de Astrofisica de Canarias. The data presented here were obtained [in part] with ALFOSC, which is provided by the Instituto de Astrofisica de Andalucia (IAA) under a joint agreement with the University of Copenhagen and NOT. VK thanks Vilho, Yrjö and Kalle Väisälä Foundation. J.J. was supported by Academy of Finland project 320085. E. L. was supported by Academy of Finland projects 317636 and 320045. Part of the French contributions is supported by the Scientific Research National Center (CNRS) and the French spatial agency (CNES). Based on observations collected at the Observatorio de Sierra Nevada, owned and operated by the Instituto de Astrof\'{i}sica de Andaluc\'{i}a (IAA-CSIC). Based on observations collected at the Centro Astron\'{o}mico Hispano-Alem\'{a}n(CAHA), proposal 22A-2.2-015, operated jointly by Junta de Andaluc\'{i}a and Consejo Superior de Investigaciones Cient\'{i}ficas (IAA-CSIC).

\bmhead{Corresponding author}
I. Liodakis, JSPS International Research Fellow, yannis.liodakis@utu.fi,

\bmhead{Author contributions}
I. Liodakis coordinated the multiwavelength observations, performed the analysis and led the writing of the paper. H. Krawczynski, A. P. Marscher, L. Peirson, P. Petrucci, J. Poutanen, F. Tavecchio, and R. Romani contributed with discussion and parts of the paper.  I. Agudo, C. Casadio, J. Escudero, and I. Myserlis contributed with the radio-millimeter polarization data. B. Ag\'{i}s-Gonz\'{a}lez, I. Agudo,  A. V. Berdyugin, M. Bernardos, G. Bonnoli, V. Casanova, M. G. Comas, C. Husillos, J. Jormanainen, V. Kravtsov, E. Lindfors, I. Liodakis, K. Nilsson, S. S. Savchenko, and A. Sota contributed with the optical polarization data. G. V. Panopoulou contributed the infrared polarization data. A. P. Marscher, G. M. Madejski, R. Middei, L. Pacciani, M. Perri, and S. Puccetti contributed the {\it Swift} and {\it NuSTAR} data. L. Di Gesu, N. di Lalla, I. Donnarumma, S. R. Ehlert, H. L. Marshall, R. Middei, M. Negro, N. Omodei, A. Paggi, and A. L. Peirson, contributed with the {\it IXPE} analysis. The remaining authors are part of the multiwavelength follow-up and {\it IXPE} teams whose significant contribution made the multiwavelength polarization observations possible.

\bmhead{Competing interests}
Authors declare that they have no competing interests.

\bmhead{Data availability}
The data that support the findings of this study are either publicly available at the HEASARC database or available from the corresponding author upon request.

\section*{Methods }

\subsection*{X-ray polarization observations}

{\it IXPE} is a joint mission of the U.S. National Aeronautics and Space Administration and the Italian Space Agency (Agenzia Spaziale Italiana).  A description of the spacecraft and of the payload is given by \cite{Weisskopf2022}; the detector units are described in \cite{Soffitta_2021}. Mrk\ 501 was observed with {\it IXPE} over an effective exposure time of 100~ksec  from 8 to 10 March 2022 (MJD 59646-59648)  and again from 26-28 March 2022 (MJD 59664-59666) for 86~ksec. The exposure times were selected based on \cite{Liodakis2019}, which determined that a 100~ksec exposure would be sufficient to measure polarization in Mrk~501 in a blind survey.  At the $\sim30''$ angular resolution of {\it IXPE}, Mrk\ 501 is essentially a point source.

The {\it IXPE}~raw (level-1) data were first reduced and corrected for instrumental polarization artifacts as well as boom and spacecraft motion to create level-2 event files (L2). The L2 data were then corrected for the energy scaling of the detector and bad aspect time intervals following standard procedures within the latest version of the {\it ixpeobssim} pipeline \cite{Pesce-Rollins2019,Baldini2022}. The {\it IXPE} L2 files contain the polarization information in the form of photon-by-photon Stokes parameters.  All the quoted results refer to the average of the three identical IXPE detector units (DU). We selected source photons using {\it xpselect} and a circular region with a radius of $60''$ centered on the source. The polarization degree and angle was determined in the 2-8~keV energy range using three different analysis techniques performed by five independent groups to ensure an unbiased estimation. Those techniques were a model-independent analysis, spectropolarimetric fit in XSPEC,  and a maximum likelihood spectropolarimetric (MLS) fit implemented within the MULTINEST algorithm. Although the effect of the photoelectric absorption is negligible over the 2-8~keV energy range of {\it IXPE}, the spectropolarimetric fits included photoelectric absorption based on the measured Galactic neutral hydrogen column density toward Mrk\ 501 of $N_H = 1.69\times10^{20}$ cm$^{-2}$ \cite{Kalberla2005}. The model-independent analysis applies the \cite{Kislat2015} formalism to a user-defined subset of photons and determines the total Stokes parameters. We have performed both a weighted and unweighted analysis. In the model-independent analysis we do not perform background subtraction. We found that the sky background counts for a $60''$ region are only 3\% of the total counts. We have verified that for a bright blazar such as Mrk~501, the background has a negligible effect on the polarization analysis. For the spectropolarimetric fits, we simultaneously fit 3$\times$ I, Q, U spectra (one set from each {\it IXPE} -- DU). In XSPEC, following the approach of \cite{Strohmayer2017}, we used an absorbed single power-law component with constant  $\Pi$ and $\psi$ (CONSTPOL model). For the MLS fit, we used a single power-law spectral component with constant intrinsic Q and U values. Given the exposure time and flux of Mrk~501 at the time of the {\it IXPE} observations, the minimum degree of detectable polarization at a 99\% confidence level (MDP99) we were able to achieve is 6.6\% for the 8-10 March, and 5.2\% for the 26-28 March observation. The source was brighter in X-rays during the 26-28 March observation (see below), hence the lower MDP99. The derived $\Pi$ and $\psi$ for the different methods are summarized in Extended Data Table \ref{tab:ixpe} for both observations. In both cases, all the measurements through the different analyses are consistent within the uncertainties with the median linear X-ray $\Pi$ and $\psi$ of $\rm\Pi_X = 10\pm2$\%, $\rm \psi_X = 134^\circ\pm5^\circ$ and $\rm\Pi_X = 11\pm2$\%, $\rm \psi_X = 115^\circ\pm4^\circ$ respectively. Extended Data Figure \ref{plt:xpol_data} shows the Stokes Q/I and Stokes U/I of our observation along with the MDP99.   Depending on the emission model, variability time scales are expected to range from sub-day to a few days \cite[e.g.,][]{DiGesu2022}. A 16-day interval between observations allows us to look for variability on a few days time scale which, however, we do not find. We have also searched for variability within the individual {\it IXPE} observations. This was done by splitting the {\it IXPE} exposures in two and three equal size time-bins. We again do not find evidence for variability within the uncertainties.

\subsection*{Multiwavelength Observations}

Here we report on a subset of our contemporaneous multiwavelength campaign from radio to TeV $\gamma$-rays which is summarized in  Extended Data Table \ref{tab:mult_obs} \& \ref{tab:mult_obs2} and Fig. \ref{plt:polssed_mrk501}. The complete multiwavelength dataset will be presented in a forthcoming paper.

\paragraph*{Millimeter-radio observations}

Polarimetric millimeter radio measurements at 3.5~mm (86.24~GHz) and 1.3~mm (230~GHz) were obtained with the 30 m Telescope of the Institut de Radioastronomie Millim\'{e}trique (IRAM), located at the Pico Veleta Observatory (Sierra Nevada, Granada, Spain), on 9-10 March 2022 (MJD 59647-59649), within the Polarimetric Monitoring of AGN at Millimeter Wavelengths (POLAMI) program (\url{http://polami.iaa.es/}, \cite{Agudo2018, Agudo2018-II, Thum2018}). Weather related reasons prevented us from obtaining radio observations during the second  {\it IXPE} exposure. Under the POLAMI observing setup, the four Stokes parameters (I, Q, U, and V) are recorded simultaneously using the XPOL procedure \cite{Thum2008}. The data reduction, calibration, and managing and flagging procedures used in POLAMI are thoroughly described in \cite{Agudo2018}.   The source was relatively stable in flux during the observations at both 1.3 and 3.5~mm with total flux densities of 0.71$\pm$0.04 Jy and 0.73$\pm$0.04 Jy at 3.5~mm, and  0.41$\pm$0.02 Jy and 0.39$\pm$0.02 Jy at 1.3~mm, on 9 and 10 of March respectively. Also, the polarized flux at 3.5~mm remained stable both in linear polarization degree and angle between the two dates. No polarization above 3.46\% (95\% confidence upper limit) was detected at 1.3~mm.

\paragraph*{Optical and infrared observations}
Optical polarization observations were performed using several telescopes across the world: the Nordic Optical Telescope (NOT) on the night of 8-9 March (MJD 59647); the Tohoku 60 cm (T60) telescope at the Haleakala Observatory on 10 March (MJD 59649) and on 28 March (MJD 59667); the 2.2m Calar Alto Observatory and 1.5m Sierra Nevada Observatory telescopes on 8-10 March; the AZT-8 telescope of the Crimean Astrophysical Observatory and the St. Petersburg State University LX-200 telescope during 8-10 March and 25-28 March.

The NOT observations used the Alhambra Faint Object Spectrograph and Camera (ALFOSC) in four bands (BVRI) in the standard polarimetric mode. The data were then analyzed with the semi-automatic pipeline developed at the Tuorla Observatory using standard photometric procedures \cite{Hovatta2016,Nilsson2018}. Both highly-polarized and unpolarized standard stars were observed during the same night for calibration purposes.  The T60 polarimetric measurements were performed using the Dipol-2 polarimeter \cite{Piirola2014}. Dipol-2 is a remotely operated double-image CCD polarimeter, which is capable of recording polarized images in three (BVR) filters simultaneously \cite{Piirola1973,Berdyugin2018,Berdyugin2019,Piirola2021}. We obtained 24 individual measurements of the Stokes Q/I and U/I parameters simultaneously in three filters (BVR). Twenty unpolarized and two highly-polarized (HD204827 and HD25443) nearby standard stars were observed for calibration and determination of the polarization angle zero point. The individual measurements were used to compute nightly average values using the ``2 $\times$ sigma-weighting algorithm''. The algorithm iteratively filters out outliers, assigning smaller weights to these measurements. The errors on the Stokes Q/I and U/I parameters were computed as standard errors of the weighted means. These errors were then used to estimate uncertainties on the polarization degree and angle \cite{Kosenkov2017,Piirola2021}. The Calar Alto Observatory observations were performed in the Johnson Cousins $\rm R_c$ optical band by the Calar Alto Faint Object Spectrograph (CAFOS) in imaging polarimetric mode on the 2.2m Telescope. The data were reduced following standard analysis procedures using both unpolarized and polarized standard stars for calibration purposes. Similarly, Mrk 501 was observed by the 1.5 m telescope at Sierra Nevada Observatory using polarized $\rm R_c$ filters during the three nights. The 70cm AZT-8 telescope and the 40cm LX-200 telescope observations were carried out in the Cousins R band. Both telescopes are equipped with nearly identical imaging photometers-polarimeters based on a ST-7 camera. Two Savart plates rotated by 45 deg relative to each other are swapped to measure the relative Stokes q and u parameters from the two split images of each source in the field. The polarization parameters for each observation are produced by the sum of 15$\times$30s consecutive exposures. The data are then corrected for bias, flat field, background level, and calibrated for instrumental and interstellar polarization using the (assumed) unpolarized comparison stars 1, 4, and 6 from \cite{Villata1998}. The same stars were used to perform differential photometry. During both {\it IXPE} observations, all the optical polarization observations are within uncertainties, which suggests no significant variability. 

Observations were also obtained with the WIRC+Pol instrument \cite{Tinyanont2019a} on the 200-inch Palomar Hale telescope in $\rm J$ band. WIRC+Pol uses a polarizing grating to disperse the light into four beams that sense the four different components of linear polarization ($0^\circ$, $45^\circ$, $90^\circ$, $135^\circ$), and a half-wave plate for beam swapping to improve polarimetric sensitivity
\citep{Tinyanont2019b,Millar-Blanchaer2021}. Data reduction made use of the WIRC+Pol Data Reduction Pipeline software ({\url{https://github.com/WIRC-Pol/wirc_drp}}, \cite{Tinyanont2019a}).  The pipeline software averages the measurements over the course of the half-wave plate rotation cycles to account for subtle differences in light paths through the instrument, and reports the degree and angle of polarization in each band. The results were verified with the use of both polarized and unpolarized standard stars. For additional details on the data reduction, see \cite{Masiero2022}.

The starlight from the host galaxy (assumed to be unpolarized) of Mrk~501 contributes a significant fraction of the optical flux. For this reason, the observed $\rm\Pi_O$ needs to be corrected for the depolarization effect of the host-galaxy. To achieve this, we need to estimate the contribution of the host galaxy ($I_\mathrm{host}$, in mJy) within the aperture used for the analysis of individual observations. The light profile of Mrk~501's host galaxy has been fully characterized in the R-band in \cite{Nilsson2007}. This allows us to estimate $I_\mathrm{host}$ for each observation separately. We then subtract $I_\mathrm{host}$ from the total intensity $I$ and estimate the intrinsic polarization degree following \cite{Hovatta2016} as $\Pi_\mathrm{intr}= \Pi_\mathrm{obs}\times{I}/(I-I_\mathrm{host})$. Due to the Dipol-2 instrument layout as well as the lack of a light profile model for the host galaxy in the J-band we are not able to accurately estimate the host-galaxy contribution to the polarization measurements for the T60 and Palomar-Hale telescopes.  For this reason, the measurements from T60 and Hale should be treated as lower limits to the intrinsic polarization degree. For the remaining telescopes, we calculate $\Pi_\mathrm{intr}$ in the R-band for each observation and then estimate a median. We find the median intrinsic polarization degree and its uncertainty to be $\rm \Pi_{\rm intr}=4\pm1\%$ for the 8-10 March observation and $\rm \Pi_{\rm intr}=5\pm1\%$ for the 26-28 March observation. Figure \ref{plt:polssed_mrk501} shows the multiwavelength polarization degree from radio to X-rays.

\paragraph*{X-ray observations}
During the {\it IXPE} observations we independently measured the X-ray total flux and spectrum with the X-Ray Telescope (XRT, \cite{Burrows2005}) on the orbiting Neil Gehrels {\it Swift} Observatory ({\it Swift}) in Window Timing mode (WT, 4$\times$1~ksec exposures -- 2$\times$1~ksec for each {\it IXPE} observation) and with the Nuclear Spectroscopic Telescope Array ({\it NuSTAR}, 20~ksec exposure,  \cite{Harrison2013}) during the 8-10 March observation. We extracted the X-ray spectrum from each telescope following standard analysis procedures and the latest calibration data files. For the source regions we used a circular radius of $47''$ and $49''$ for {\it Swift} and {\it NuSTAR}, respectively. To estimate the background for the {\it NuSTAR} spectra we used a $147''$ circular region outside of the region containing significant photon counts from Mrk 501. The background for {\it Swift} was extracted using the same size circular region from an available blank sky WT observation from the {\it Swift archive}. For the 8-10 March observation, we fit the combined {\it Swift} and {\it NuSTAR} data in {\it XSPEC} with an absorbed log-parabola model $N(E) = (E/E_p)^{(-\alpha-\beta\log(E/E_p))}$, in the 0.3-79~keV energy range. $\rm N_H$ was set to the Galactic value, and the pivot energy was set to $E_{p}=$5~keV. This model provides a reasonably good fit to the data ($\rm \chi^2/dof=862/850$) with best-fit parameters $\alpha=2.27\pm0.01$ and $\beta=0.28\pm0.01$. We also tested a single power-law model, however, there is clear curvature in the spectrum and the fit is statistically worse ($\rm \chi^2/dof=2005/851$). We measure the flux of the source in the 2-8~keV range to be (10.0$\pm$0.5)$\rm \times10^{-11}~erg/s/cm^2$. We do not find evidence for variability during the {\it IXPE} observations. For the 26-28 March observation we follow the same procedure using only the available {\it Swift} data. The source was in a higher flux state with  $\alpha=2.05\pm0.02$ and $\beta=0.26\pm0.04$ and flux in the 2-8~keV range of (21.0$\pm$0.6)$\rm \times10^{-11}~erg/s/cm^2$.  The {\it Swift} observations show a change from 12 to 14 counts/sec (17\% increase) from the beginning until the end of the {\it IXPE} observation. The results from our multiwavelength campaign are summarized in Extended Data Table \ref{tab:mult_obs} and \ref{tab:mult_obs2}.

\subsection*{Activity state of Mrk~501}

Mrk\ 501 is a BL Lac object at a redshift of $z=0.033$, corresponding to a luminosity distance of 141.3~Mpc, assuming a flat $\Lambda$CDM cosmological model with $\Omega_m$=0.27 and H$_0$=71~km/s/Mpc \cite{Komatsu2009}, and a synchrotron peak frequency $\rm \nu_{\rm syn}\sim2.8\times10^{15}~Hz$ \cite{Ajello2020}. It is among the brightest sources in the sky at very high $\gamma$-ray energies ($\geq0.1$ Tev), and is well-studied across the electromagnetic spectrum \cite[e.g.,][]{Quinn1996,Abdo2011,Aleksic2015,Ahnen2017,Arbet-Engels2021-II}. We use archival data from {\it Swift} (\url{https://www.swift.psu.edu/monitoring/}),  Steward observatory (\url{http://james.as.arizona.edu/~psmith/Fermi/},\cite{Smith2009}), the RoboPol program (\url{http://robopol.physics.uoc.gr/}, \cite{Blinov2021}), and the Boston University (BU) blazar monitoring program (\url{https://www.bu.edu/blazars/index.html}) to build the long-term light curves of  Mrk~501 in  optical  brightness (R-band magnitude), optical polarization degree, polarization angle, and X-ray flux (Fig. \ref{plt:light_mrk501}). The optical observations cover a range from October 2008 up to June 2021. At R band, the source varied between $13.53^m$ and $13.24^m$, with a median of $13.4^m$. The median observed $\rm\Pi_O$ (not corrected for the host-galaxy contribution) was 2.1\% with a minimum of 0.07\% and a maximum of 5.9\%. The $\psi_O$ typically fluctuates about the jet axis (120$^\circ\pm12^\circ$) with a median of 136$^\circ$, and a minimum and maximum of 65$^\circ$ and 171$^\circ$ respectively. The X-ray observations cover a range from April 2005 until June 2020. The median X-ray flux in the 0.3-10~keV was 15$\rm\times10^{-11} erg/s/cm^2$, with a minimum and maximum at around 3.7$\rm\times10^{-11} erg/s/cm^2$ and 76$\rm\times10^{-11} erg/s/cm^2$, respectively. At the time of the {\it IXPE} observations our multiwavelength campaign finds the flux and polarization of the source within one standard deviation of the median of the respective archival light curve. For the first {\it IXPE} observation the X-ray flux of the source seems to correspond to an average state, while in the second observation we find the source in a slightly elevated-flux state.\\

\newpage
\renewcommand{\figurename}{Extended Data Table} 
\setcounter{figure}{0}    
\begin{figure}
\centering
\includegraphics[scale=0.25]{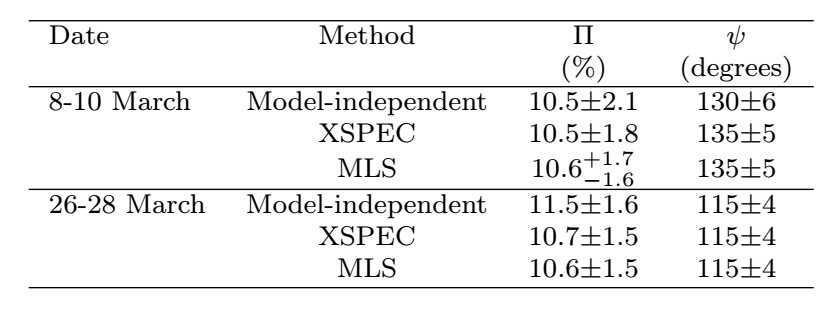}
 \caption{Median polarization degree and angle measurements from the {\it IXPE} data analysis performed by independent groups using three analysis techniques.}
\label{tab:ixpe}
\end{figure}

\renewcommand{\figurename}{Extended Data Table} 
\setcounter{figure}{1}    
\begin{figure}
\centering
\includegraphics[scale=0.3]{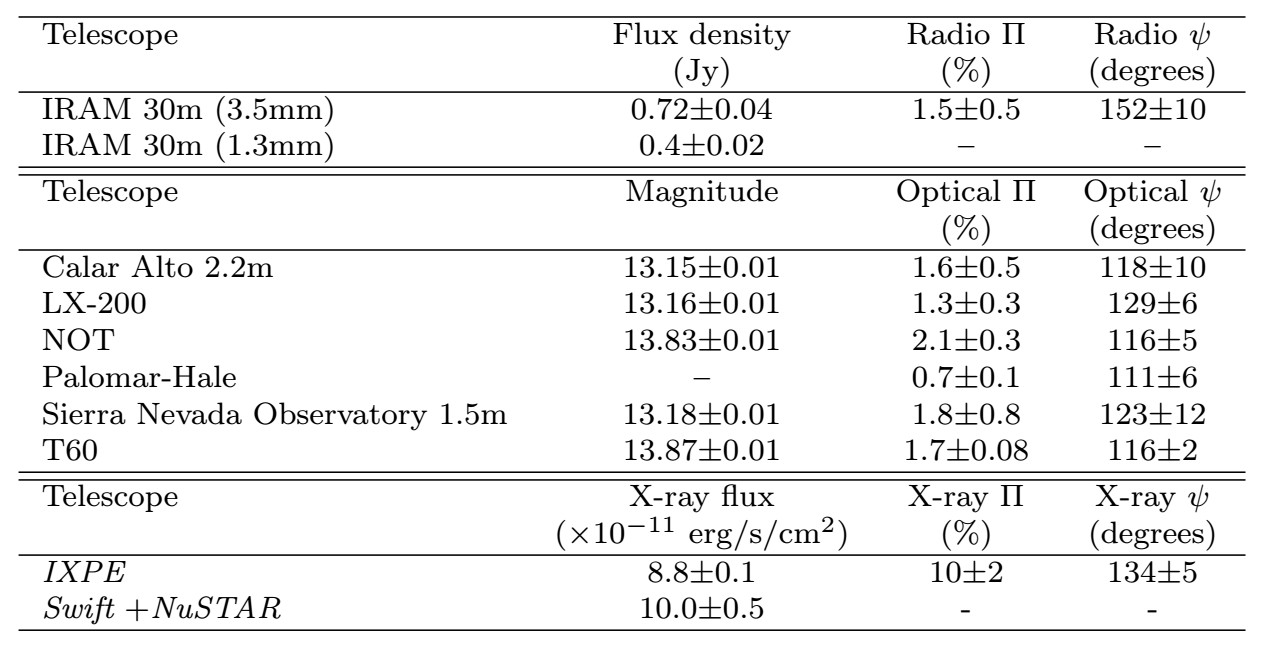}
\caption{Multiwavelenth and polarization observations for the 2022 March 8-10 observation.}
\label{tab:mult_obs}
\end{figure}
\noindent{ {\it Table comments:} \small The millimeter-radio flux density is in Janskys. For the millimeter-radio and optical observations we report the median estimate of the observations during the {\it IXPE} observation. The listed uncertainty is either the standard deviation of the measurements or the median uncertainty, whichever is larger. For the NOT and T60 analysis we used a circular $1.5''$ radius aperture. For the data analysis of remaining optical telescopes we used a $7.5''$ aperture. The Palomar observations are in the J-band. $\psi$ is given in degrees. The X-ray fluxes are estimated in the 2-8~keV range, and given in units of $\rm 10^{-11}~erg/s/cm^2$.}

\newpage
\renewcommand{\figurename}{Extended Data Table} 
\setcounter{figure}{2}    
\begin{figure}[ht]
\centering
\includegraphics[scale=0.25]{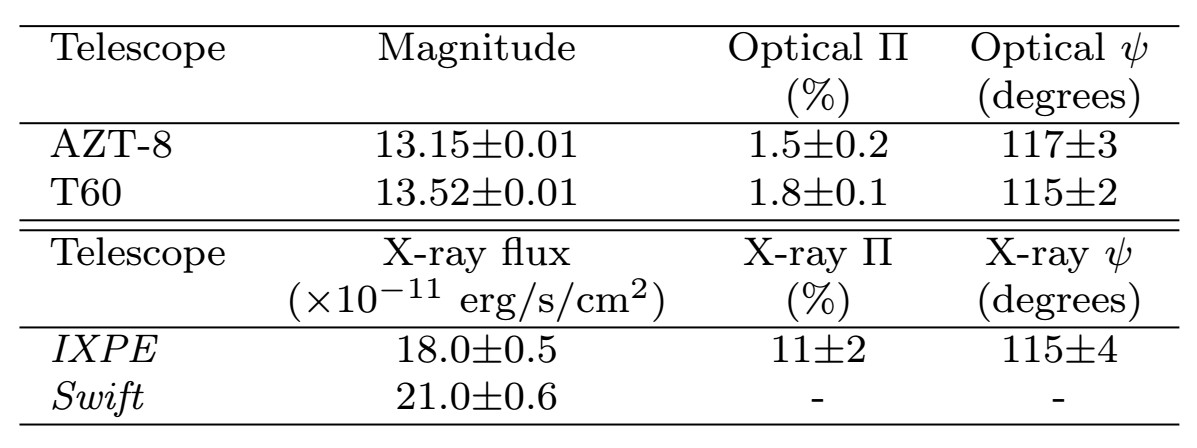}
 \caption{Multiwavelenth and polarization observations for the 2022 March 26-28 observation.}
\label{tab:mult_obs2}
\end{figure}
\noindent{{\it Table comments:} \small Same as in Extended Data Table \ref{tab:mult_obs}.}

\renewcommand{\figurename}{Extended Data Figure} 
\setcounter{figure}{0}    
\begin{figure}
\centering
\includegraphics[scale=0.25]{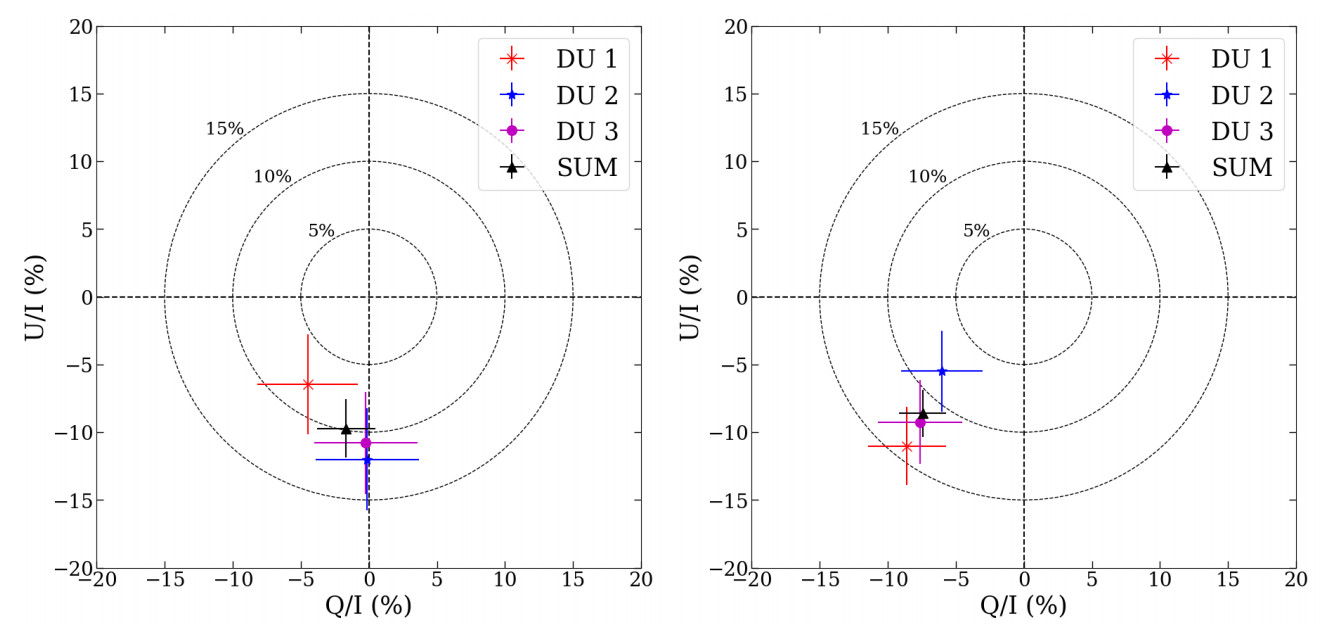}
 \caption{Stokes Q/I and Stokes U/I parameters of our {\it IXPE} observations during 8-10 March 2022 (left) and 26-28 March 2022 (right). The measurements are shown for the three detectors (DU1 [red x], DU2 [blue star], DU3 [magenta circle]) separately and combined (black triangle). In both panels error bars denote the 68\% CI}
\label{plt:xpol_data}
\end{figure}


\begin{thebibliography}{10}
\expandafter\ifx\csname url\endcsname\relax
  \def\url#1{\burl{#1}}\fi
\expandafter\ifx\csname urlprefix\endcsname\relax\def\urlprefix{URL }\fi
\providecommand{\bibinfo}[2]{#2}
\providecommand{\eprint}[2][]{\url{#2}}
\providecommand{\doi}[1]{\url{https://doi.org/#1}}
\bibcommenthead

\bibitem{Jorstad2005}
\bibinfo{author}{{Jorstad}, S.~G.} \emph{et~al.}
\newblock \bibinfo{title}{{Polarimetric Observations of 15 Active Galactic
  Nuclei at High Frequencies: Jet Kinematics from Bimonthly Monitoring with the
  Very Long Baseline Array}}.
\newblock \emph{\bibinfo{journal}{AJ}} \textbf{\bibinfo{volume}{130}},
  \bibinfo{pages}{1418--1465} (\bibinfo{year}{2005}).

  .

\bibitem{Marin2018}
\bibinfo{author}{{Marin}, F.}
\newblock \bibinfo{title}{{A complete disclosure of the hidden type-1 AGN in
  NGC 1068 thanks to 52 yr of broad-band polarimetric observation}}.
\newblock \emph{\bibinfo{journal}{\mnras}} \textbf{\bibinfo{volume}{479}}~(3),
  \bibinfo{pages}{3142--3154} (\bibinfo{year}{2018}).



\bibitem{Blinov2021}
\bibinfo{author}{{Blinov}, D.} \emph{et~al.}
\newblock \bibinfo{title}{{RoboPol: AGN polarimetric monitoring data}}.
\newblock \emph{\bibinfo{journal}{\mnras}} \textbf{\bibinfo{volume}{501}}~(3),
  \bibinfo{pages}{3715--3726} (\bibinfo{year}{2021}).


\bibitem{Marscher1985}
\bibinfo{author}{{Marscher}, A.~P.} \& \bibinfo{author}{{Gear}, W.~K.}
\newblock \bibinfo{title}{{Models for high-frequency radio outbursts in
  extragalactic sources, with application to the early 1983
  millimeter-to-infrared flare of 3C 273}}.
\newblock \emph{\bibinfo{journal}{\apj}} \textbf{\bibinfo{volume}{298}},
  \bibinfo{pages}{114--127} (\bibinfo{year}{1985}).


\bibitem{Marscher2014}
\bibinfo{author}{{Marscher}, A.~P.}
\newblock \bibinfo{title}{{Turbulent, Extreme Multi-zone Model for Simulating
  Flux and Polarization Variability in Blazars}}.
\newblock \emph{\bibinfo{journal}{\apj}} \textbf{\bibinfo{volume}{780}},
  \bibinfo{pages}{87} (\bibinfo{year}{2014}).
 
 
\bibitem{Tavecchio2018}
\bibinfo{author}{{Tavecchio}, F.}, \bibinfo{author}{{Landoni}, M.},
  \bibinfo{author}{{Sironi}, L.} \& \bibinfo{author}{{Coppi}, P.}
\newblock \bibinfo{title}{{Probing dissipation mechanisms in BL Lac jets
  through X-ray polarimetry}}.
\newblock \emph{\bibinfo{journal}{\mnras}} \textbf{\bibinfo{volume}{480}}~(3),
  \bibinfo{pages}{2872--2880} (\bibinfo{year}{2018}).    
  
  
\bibitem{Weisskopf2022}
\bibinfo{author}{Weisskopf, M.~C.} \emph{et~al.}
\newblock \bibinfo{title}{{Imaging X-ray Polarimetry Explorer: prelaunch}}.
\newblock \emph{\bibinfo{journal}{Journal of Astronomical Telescopes,
  Instruments, and Systems}} \textbf{\bibinfo{volume}{8}}~(2),
  \bibinfo{pages}{1 -- 28} (\bibinfo{year}{2022}).  
 
\bibitem{Peirson2018}
\bibinfo{author}{{Peirson}, A.~L.} \& \bibinfo{author}{{Romani}, R.~W.}
\newblock \bibinfo{title}{{The Polarization Behavior of Relativistic
  Synchrotron Jets}}.
\newblock \emph{\bibinfo{journal}{\apj}} \textbf{\bibinfo{volume}{864}},
  \bibinfo{pages}{140} (\bibinfo{year}{2018}).

   

   

\bibitem{Peirson2019}
\bibinfo{author}{{Peirson}, A.~L.} \& \bibinfo{author}{{Romani}, R.~W.}
\newblock \bibinfo{title}{{The Polarization Behavior of Relativistic
  Synchrotron Self-Compton Jets}}.
\newblock \emph{\bibinfo{journal}{\apj}} \textbf{\bibinfo{volume}{885}}~(1),
  \bibinfo{pages}{76} (\bibinfo{year}{2019}).




\bibitem{MJ2021}
\bibinfo{author}{{Marscher}, A.~P.} \& \bibinfo{author}{{Jorstad}, S.~G.}
\newblock \bibinfo{title}{{Frequency and Time Dependence of Linear Polarization
  in Turbulent Jets of Blazars}}.
\newblock \emph{\bibinfo{journal}{Galaxies}} \textbf{\bibinfo{volume}{9}}~(2),
  \bibinfo{pages}{27} (\bibinfo{year}{2021}).




\bibitem{Marscher2008}
\bibinfo{author}{{Marscher}, A.~P.} \emph{et~al.}
\newblock \bibinfo{title}{{The inner jet of an active galactic nucleus as
  revealed by a radio-to-{$\gamma$}-ray outburst}}.
\newblock \emph{\bibinfo{journal}{\nat}} \textbf{\bibinfo{volume}{452}},
  \bibinfo{pages}{966--969} (\bibinfo{year}{2008}).



\bibitem{Kiehlmann2017}
\bibinfo{author}{{Kiehlmann}, S.}, \bibinfo{author}{{Blinov}, D.},
  \bibinfo{author}{{Pearson}, T.~J.} \& \bibinfo{author}{{Liodakis}, I.}
\newblock \bibinfo{title}{{Optical EVPA rotations in blazars: testing a
  stochastic variability model with RoboPol data}}.
\newblock \emph{\bibinfo{journal}{\mnras}} \textbf{\bibinfo{volume}{472}},
  \bibinfo{pages}{3589--3604} (\bibinfo{year}{2017}).


\bibitem{Blinov2018}
\bibinfo{author}{{Blinov}, D.} \emph{et~al.}
\newblock \bibinfo{title}{{RoboPol: connection between optical polarization
  plane rotations and gamma-ray flares in blazars}}.
\newblock \emph{\bibinfo{journal}{\mnras}} \textbf{\bibinfo{volume}{474}},
  \bibinfo{pages}{1296--1306} (\bibinfo{year}{2018}).


\bibitem{Tavecchio2020}
\bibinfo{author}{{Tavecchio}, F.}, \bibinfo{author}{{Landoni}, M.},
  \bibinfo{author}{{Sironi}, L.} \& \bibinfo{author}{{Coppi}, P.}
\newblock \bibinfo{title}{{Probing shock acceleration in BL Lac jets through
  X-ray polarimetry: the time-dependent view}}.
\newblock \emph{\bibinfo{journal}{\mnras}} \textbf{\bibinfo{volume}{498}}~(1),
  \bibinfo{pages}{599--608} (\bibinfo{year}{2020}).


\bibitem{Lyutikov2005}
\bibinfo{author}{{Lyutikov}, M.}, \bibinfo{author}{{Pariev}, V.~I.} \&
  \bibinfo{author}{{Gabuzda}, D.~C.}
\newblock \bibinfo{title}{{Polarization and structure of relativistic
  parsec-scale AGN jets}}.
\newblock \emph{\bibinfo{journal}{\mnras}} \textbf{\bibinfo{volume}{360}}~(3),
  \bibinfo{pages}{869--891} (\bibinfo{year}{2005}).


\bibitem{Hovatta2012}
\bibinfo{author}{{Hovatta}, T.} \emph{et~al.}
\newblock \bibinfo{title}{{MOJAVE: Monitoring of Jets in Active Galactic Nuclei
  with VLBA Experiments. VIII. Faraday Rotation in Parsec-scale AGN Jets}}.
\newblock \emph{\bibinfo{journal}{\aj}} \textbf{\bibinfo{volume}{144}}~(4),
  \bibinfo{pages}{105} (\bibinfo{year}{2012}).


\bibitem{Gabuzda2021}
\bibinfo{author}{{Gabuzda}, D.~C.}
\newblock \bibinfo{title}{{Inherent and Local Magnetic Field Structures in Jets
  from Active Galactic Nuclei}}.
\newblock \emph{\bibinfo{journal}{Galaxies}} \textbf{\bibinfo{volume}{9}}~(3),
  \bibinfo{pages}{58} (\bibinfo{year}{2021}).







\bibitem{DiGesu2022}
\bibinfo{author}{{Di Gesu}, L.} \emph{et~al.}
\newblock \bibinfo{title}{{Testing particle acceleration models for BL LAC jets
  with the Imaging X-ray Polarimetry Explorer}}.
\newblock \emph{\bibinfo{journal}{\aap}} \textbf{\bibinfo{volume}{662}},
  \bibinfo{pages}{A83} (\bibinfo{year}{2022}).







\bibitem{Angelakis2016}
\bibinfo{author}{{Angelakis}, E.} \emph{et~al.}
\newblock \bibinfo{title}{{RoboPol: the optical polarization of gamma-ray-loud
  and gamma-ray-quiet blazars}}.
\newblock \emph{\bibinfo{journal}{\mnras}} \textbf{\bibinfo{volume}{463}}~(3),
  \bibinfo{pages}{3365--3380} (\bibinfo{year}{2016}).







\bibitem{Sironi2021}
\bibinfo{author}{{Sironi}, L.}, \bibinfo{author}{{Rowan}, M.~E.} \&
  \bibinfo{author}{{Narayan}, R.}
\newblock \bibinfo{title}{{Reconnection-driven Particle Acceleration in
  Relativistic Shear Flows}}.
\newblock \emph{\bibinfo{journal}{\apjl}} \textbf{\bibinfo{volume}{907}}~(2),
  \bibinfo{pages}{L44} (\bibinfo{year}{2021}).


\bibitem{Tavecchio2021}
\bibinfo{author}{{Tavecchio}, F.}
\newblock \bibinfo{title}{{Probing Magnetic Fields and Acceleration Mechanisms
  in Blazar Jets with X-ray Polarimetry}}.
\newblock \emph{\bibinfo{journal}{Galaxies}} \textbf{\bibinfo{volume}{9}}~(2),
  \bibinfo{pages}{37} (\bibinfo{year}{2021}).





\bibitem{Weaver2022}
\bibinfo{author}{{Weaver}, Z.~R.} \emph{et~al.}
\newblock \bibinfo{title}{{Kinematics of Parsec-Scale Jets of Gamma-Ray Bright
  Blazars at 43 GHz during Ten Years of the VLBA-BU-BLAZAR Program}}.
\newblock \emph{\bibinfo{journal}{\apjs}} \textbf{\bibinfo{volume}{260}},
  \bibinfo{pages}{12} (\bibinfo{year}{2022}).





\bibitem{Webb2021}
\bibinfo{author}{{Webb}, J.~R.} \emph{et~al.}
\newblock \bibinfo{title}{{The Nature of Micro-Variability in Blazars}}.
\newblock \emph{\bibinfo{journal}{Galaxies}} \textbf{\bibinfo{volume}{9}}~(4),
  \bibinfo{pages}{114} (\bibinfo{year}{2021}).


\bibitem{Blinov2016}
\bibinfo{author}{{Blinov}, D.} \emph{et~al.}
\newblock \bibinfo{title}{{RoboPol: do optical polarization rotations occur in
  all blazars?}}
\newblock \emph{\bibinfo{journal}{\mnras}} \textbf{\bibinfo{volume}{462}}~(2),
  \bibinfo{pages}{1775--1785} (\bibinfo{year}{2016}).





\section*{\Large References}



\bibitem{Soffitta_2021}
P.~Soffitta, L.~Baldini, R.~Bellazzini, E.~Costa, L.~Latronico, F.~Muleri, E.D.
  Monte, S.~Fabiani, M.~Minuti, M.~Pinchera, C.~Sgro', G.~Spandre, A.~Trois,
  F.~Amici, H.~Andersson, P.~Attina', M.~Bachetti, M.~Barbanera, F.~Borotto,
  A.~Brez, D.~Brienza, C.~Caporale, C.~Cardelli, R.~Carpentiero, S.~Castellano,
  M.~Castronuovo, L.~Cavalli, E.~Cavazzuti, M.~Ceccanti, M.~Centrone,
  S.~Ciprini, S.~Citraro, F.~D'Amico, E.~D'Alba, S.D. Cosimo, N.D. Lalla, A.D.
  Marco, G.D. Persio, I.~Donnarumma, Y.~Evangelista, R.~Ferrazzoli, A.~Hayato,
  T.~Kitaguchi, F.L. Monaca, C.~Lefevre, P.~Loffredo, P.~Lorenzi, L.~Lucchesi,
  C.~Magazzu, S.~Maldera, A.~Manfreda, E.~Mangraviti, M.~Marengo, G.~Matt,
  P.~Mereu, A.~Morbidini, F.~Mosti, T.~Nakano, H.~Nasimi, B.~Negri, S.~Nenonen,
  A.~Nuti, L.~Orsini, M.~Perri, M.~Pesce-Rollins, R.~Piazzolla, M.~Pilia,
  A.~Profeti, S.~Puccetti, J.~Rankin, A.~Ratheesh, A.~Rubini, F.~Santoli,
  P.~Sarra, E.~Scalise, A.~Sciortino, T.~Tamagawa, M.~Tardiola, A.~Tobia,
  M.~Vimercati, F.~Xie, The instrument of the imaging x-ray polarimetry
  explorer.
\newblock The Astronomical Journal \textbf{162}(5), 208 (2021).


\bibitem{Liodakis2019}
\bibinfo{author}{{Liodakis}, I.}, \bibinfo{author}{{Peirson}, A.~L.} \&
  \bibinfo{author}{{Romani}, R.~W.}
\newblock \bibinfo{title}{{Prospects for Detecting X-ray Polarization in Blazar Jets}}.
\newblock \emph{\bibinfo{journal}{The Astrophysical Journal}} \textbf{\bibinfo{volume}{880}}, \bibinfo{pages}{7}
  (\bibinfo{year}{2019}).


\bibitem{Pesce-Rollins2019}
\bibinfo{author}{{Pesce-Rollins}, M.}, \bibinfo{author}{{Lalla}, N.~D.},
  \bibinfo{author}{{Omodei}, N.} \& \bibinfo{author}{{Baldini}, L.}
\newblock \bibinfo{title}{{An observation-simulation and analysis framework for
  the Imaging X-ray Polarimetry Explorer (IXPE)}}.
\newblock \emph{\bibinfo{journal}{Nuclear Instruments and Methods in Physics
  Research A}} \textbf{\bibinfo{volume}{936}}, \bibinfo{pages}{224--226}
  (\bibinfo{year}{2019}).


\bibitem{Baldini2022}
\bibinfo{author}{{Baldini}, L.} \emph{et~al.}
\newblock \bibinfo{title}{{ixpeobssim: a Simulation and Analysis Framework for
  the Imaging X-ray Polarimetry Explorer}}.
\newblock \emph{\bibinfo{journal}{arXiv e-prints}}
  \bibinfo{pages}{arXiv:2203.06384} (\bibinfo{year}{2022}).


      
      
\bibitem{Kalberla2005}
\bibinfo{author}{{Kalberla}, P.~M.~W.} \emph{et~al.}
\newblock \bibinfo{title}{{The Leiden/Argentine/Bonn (LAB) Survey of Galactic
  HI. Final data release of the combined LDS and IAR surveys with improved
  stray-radiation corrections}}.
\newblock \emph{\bibinfo{journal}{\aap}} \textbf{\bibinfo{volume}{440}}~(2),
  \bibinfo{pages}{775--782} (\bibinfo{year}{2005}).


\bibitem{Kislat2015}
\bibinfo{author}{{Kislat}, F.}, \bibinfo{author}{{Clark}, B.},
  \bibinfo{author}{{Beilicke}, M.} \& \bibinfo{author}{{Krawczynski}, H.}
\newblock \bibinfo{title}{{Analyzing the data from X-ray polarimeters with
  Stokes parameters}}.
\newblock \emph{\bibinfo{journal}{Astroparticle Physics}}
  \textbf{\bibinfo{volume}{68}}, \bibinfo{pages}{45--51}
  (\bibinfo{year}{2015}).

\bibitem{Strohmayer2017}
\bibinfo{author}{{Strohmayer}, T.~E.}
\newblock \bibinfo{title}{{X-Ray Spectro-polarimetry with Photoelectric Polarimeters}}.
\newblock \emph{\bibinfo{journal}{\apj}} \textbf{\bibinfo{volume}{838}}~(1),
  \bibinfo{pages}{72} (\bibinfo{year}{2017}).


\bibitem{Agudo2018}
\bibinfo{author}{{Agudo}, I.} \emph{et~al.}
\newblock \bibinfo{title}{{POLAMI: Polarimetric Monitoring of AGN at Millimetre
  Wavelengths - I. The programme, calibration and calibrator data products}}.
\newblock \emph{\bibinfo{journal}{\mnras}} \textbf{\bibinfo{volume}{474}}~(2),
  \bibinfo{pages}{1427--1435} (\bibinfo{year}{2018}).


\bibitem{Agudo2018-II}
\bibinfo{author}{{Agudo}, I.} \emph{et~al.}
\newblock \bibinfo{title}{{POLAMI: Polarimetric Monitoring of Active Galactic
  Nuclei at Millimetre Wavelengths - III. Characterization of total flux
  density and polarization variability of relativistic jets}}.
\newblock \emph{\bibinfo{journal}{\mnras}} \textbf{\bibinfo{volume}{473}}~(2),
  \bibinfo{pages}{1850--1867} (\bibinfo{year}{2018}).


\bibitem{Thum2018}
\bibinfo{author}{{Thum}, C.} \emph{et~al.}
\newblock \bibinfo{title}{{POLAMI: Polarimetric Monitoring of Active Galactic
  Nuclei at Millimetre Wavelengths - II. Widespread circular polarization}}.
\newblock \emph{\bibinfo{journal}{\mnras}} \textbf{\bibinfo{volume}{473}}~(2),
  \bibinfo{pages}{2506--2520} (\bibinfo{year}{2018}).


\bibitem{Thum2008}
\bibinfo{author}{{Thum}, C.}, \bibinfo{author}{{Wiesemeyer}, H.},
  \bibinfo{author}{{Paubert}, G.}, \bibinfo{author}{{Navarro}, S.} \&
  \bibinfo{author}{{Morris}, D.}
\newblock \bibinfo{title}{{XPOL{\textemdash}the Correlation Polarimeter at the
  IRAM 30-m Telescope}}.
\newblock \emph{\bibinfo{journal}{\pasp}} \textbf{\bibinfo{volume}{120}}~(869),
  \bibinfo{pages}{777} (\bibinfo{year}{2008}).


\bibitem{Hovatta2016}
\bibinfo{author}{{Hovatta}, T.} \emph{et~al.}
\newblock \bibinfo{title}{{Optical polarization of high-energy BL Lacertae
  objects}}.
\newblock \emph{\bibinfo{journal}{\aap}} \textbf{\bibinfo{volume}{596}},
  \bibinfo{pages}{A78} (\bibinfo{year}{2016}).


\bibitem{Nilsson2018}
\bibinfo{author}{{Nilsson}, K.} \emph{et~al.}
\newblock \bibinfo{title}{{Long-term optical monitoring of TeV emitting
  blazars. I. Data analysis}}.
\newblock \emph{\bibinfo{journal}{\aap}} \textbf{\bibinfo{volume}{620}},
  \bibinfo{pages}{A185} (\bibinfo{year}{2018}).


\bibitem{Piirola2014}
\bibinfo{author}{{Piirola}, V.}, \bibinfo{author}{{Berdyugin}, A.} \&
  \bibinfo{author}{{Berdyugina}, S.}
\newblock \bibinfo{editor}{{Ramsay}, S.~K.}, \bibinfo{editor}{{McLean}, I.~S.}
  \& \bibinfo{editor}{{Takami}, H.} (eds) \emph{\bibinfo{title}{{DIPOL-2: a
  double image high precision polarimeter}}}.
\newblock (eds \bibinfo{editor}{{Ramsay}, S.~K.}, \bibinfo{editor}{{McLean},
  I.~S.} \& \bibinfo{editor}{{Takami}, H.})
  \emph{\bibinfo{booktitle}{Ground-based and Airborne Instrumentation for
  Astronomy V}}, Vol. \bibinfo{volume}{9147} of \emph{\bibinfo{series}{Society
  of Photo-Optical Instrumentation Engineers (SPIE) Conference Series}},
  \bibinfo{pages}{91478I} (\bibinfo{year}{2014}).

\bibitem{Piirola1973}
\bibinfo{author}{{Piirola}, V.}
\newblock \bibinfo{title}{{A double image chopping polarimeter.}}
\newblock \emph{\bibinfo{journal}{\aap}} \textbf{\bibinfo{volume}{27}},
  \bibinfo{pages}{383--388} (\bibinfo{year}{1973}) .

\bibitem{Berdyugin2018}
\bibinfo{author}{{Berdyugin}, A.~V.}, \bibinfo{author}{{Berdyugina}, S.~V.} \&
  \bibinfo{author}{{Piirola}, V.}
\newblock \bibinfo{editor}{{Evans}, C.~J.}, \bibinfo{editor}{{Simard}, L.} \&
  \bibinfo{editor}{{Takami}, H.} (eds) \emph{\bibinfo{title}{{High-precision
  and high-accuracy polarimetry of exoplanets}}}.
\newblock (eds \bibinfo{editor}{{Evans}, C.~J.}, \bibinfo{editor}{{Simard}, L.}
  \& \bibinfo{editor}{{Takami}, H.}) \emph{\bibinfo{booktitle}{Ground-based and
  Airborne Instrumentation for Astronomy VII}}, Vol. \bibinfo{volume}{10702} of
  \emph{\bibinfo{series}{Society of Photo-Optical Instrumentation Engineers
  (SPIE) Conference Series}}, \bibinfo{pages}{107024Z} (\bibinfo{year}{2018}).

\bibitem{Berdyugin2019}
\bibinfo{author}{{Berdyugin}, A.}, \bibinfo{author}{{Piirola}, V.} \&
  \bibinfo{author}{{Poutanen}, J.}
\newblock \bibinfo{editor}{{Mignani}, R.}, \bibinfo{editor}{{Shearer}, A.},
  \bibinfo{editor}{{S{\l}owikowska}, A.} \& \bibinfo{editor}{{Zane}, S.} (eds)
  \emph{\bibinfo{title}{{Optical Polarimetry: Methods, Instruments and
  Calibration Techniques}}}.
\newblock (eds \bibinfo{editor}{{Mignani}, R.}, \bibinfo{editor}{{Shearer},
  A.}, \bibinfo{editor}{{S{\l}owikowska}, A.} \& \bibinfo{editor}{{Zane}, S.})
  \emph{\bibinfo{booktitle}{Astronomical Polarisation from the Infrared to
  Gamma Rays}}, Vol. \bibinfo{volume}{460} of
  \emph{\bibinfo{series}{Astrophysics and Space Science Library}},
  \bibinfo{pages}{33} (\bibinfo{year}{2019}).


\bibitem{Piirola2021}
\bibinfo{author}{{Piirola}, V.}, \bibinfo{author}{{Kosenkov}, I.~A.},
  \bibinfo{author}{{Berdyugin}, A.~V.}, \bibinfo{author}{{Berdyugina}, S.~V.}
  \& \bibinfo{author}{{Poutanen}, J.}
\newblock \bibinfo{title}{{Double Image Polarimeter{\textemdash}Ultra Fast:
  Simultaneous Three-color (BV R) Polarimeter with Electron-multiplying
  Charge-coupled Devices}}.
\newblock \emph{\bibinfo{journal}{\aj}} \textbf{\bibinfo{volume}{161}}~(1),
  \bibinfo{pages}{20} (\bibinfo{year}{2021}).


\bibitem{Kosenkov2017}
\bibinfo{author}{{Kosenkov}, I.~A.} \emph{et~al.}
\newblock \bibinfo{title}{{High-precision optical polarimetry of the accreting
  black hole V404 Cyg during the 2015 June outburst}}.
\newblock \emph{\bibinfo{journal}{\mnras}} \textbf{\bibinfo{volume}{468}}~(4),
  \bibinfo{pages}{4362--4373} (\bibinfo{year}{2017}).


\bibitem{Villata1998}
\bibinfo{author}{{Villata}, M.}, \bibinfo{author}{{Raiteri}, C.~M.},
  \bibinfo{author}{{Lanteri}, L.}, \bibinfo{author}{{Sobrito}, G.} \&
  \bibinfo{author}{{Cavallone}, M.}
\newblock \bibinfo{title}{{BVR photometry of comparison stars in selected
  blazar fields. I. Photometric sequences for 10 BL Lacertae objects}}.
\newblock \emph{\bibinfo{journal}{\aaps}} \textbf{\bibinfo{volume}{130}},
  \bibinfo{pages}{305--310} (\bibinfo{year}{1998}).


\bibitem{Tinyanont2019a}
\bibinfo{author}{{Tinyanont}, S.} \emph{et~al.}
\newblock \bibinfo{title}{{WIRC+Pol: A Low-resolution Near-infrared
  Spectropolarimeter}}.
\newblock \emph{\bibinfo{journal}{\pasp}} \textbf{\bibinfo{volume}{131}}~(996),
  \bibinfo{pages}{025001} (\bibinfo{year}{2019}).


\bibitem{Tinyanont2019b}
\bibinfo{author}{{Tinyanont}, S.} \emph{et~al.}
\newblock \emph{\bibinfo{title}{{Achieving a spectropolarimetric precision
  better than 0.1\% in the near-infrared with WIRC+Pol}}}, Vol.
  \bibinfo{volume}{11132} of \emph{\bibinfo{series}{Society of Photo-Optical
  Instrumentation Engineers (SPIE) Conference Series}},
  \bibinfo{pages}{1113209} (\bibinfo{year}{2019}).


\bibitem{Millar-Blanchaer2021}
\bibinfo{author}{Millar-Blanchaer, M.~A.} \emph{et~al.}
\newblock \bibinfo{editor}{Evans, C.~J.}, \bibinfo{editor}{Bryant, J.~J.} \&
  \bibinfo{editor}{Motohara, K.} (eds) \emph{\bibinfo{title}{{WIRC-POL: 0.1%
  polarimetric precision on-sky with the installation of a HWP modulator}}}.
\newblock (eds \bibinfo{editor}{Evans, C.~J.}, \bibinfo{editor}{Bryant, J.~J.}
  \& \bibinfo{editor}{Motohara, K.}) \emph{\bibinfo{booktitle}{Ground-based and
  Airborne Instrumentation for Astronomy VIII}}, Vol. \bibinfo{volume}{11447},
  \bibinfo{pages}{1315 -- 1332}. \bibinfo{organization}{International Society
  for Optics and Photonics} (\bibinfo{publisher}{SPIE}, \bibinfo{year}{2021}).


\bibitem{Masiero2022}
\bibinfo{author}{{Masiero}, J.~R.}, \bibinfo{author}{{Tinyanont}, S.} \&
  \bibinfo{author}{{Millar-Blanchaer}, M.~A.}
\newblock \bibinfo{title}{{Asteroid Polarimetric-Phase Behavior in the
  Near-Infrared: S- and C-Complex Objects}}.
\newblock \emph{\bibinfo{journal}{arXiv e-prints}}
  \bibinfo{pages}{arXiv:2203.15790} (\bibinfo{year}{2022}).


\bibitem{Nilsson2007}
\bibinfo{author}{{Nilsson}, K.} \emph{et~al.}
\newblock \bibinfo{title}{{Host galaxy subtraction of TeV candidate BL Lacertae
  objects}}.
\newblock \emph{\bibinfo{journal}{\aap}} \textbf{\bibinfo{volume}{475}}~(1),
  \bibinfo{pages}{199--207} (\bibinfo{year}{2007}).


\bibitem{Burrows2005}
\bibinfo{author}{{Burrows}, D.~N.} \emph{et~al.}
\newblock \bibinfo{title}{{The Swift X-Ray Telescope}}.
\newblock \emph{\bibinfo{journal}{\ssr}} \textbf{\bibinfo{volume}{120}}~(3-4),
  \bibinfo{pages}{165--195} (\bibinfo{year}{2005}).


\bibitem{Harrison2013}
\bibinfo{author}{{Harrison}, F.~A.} \emph{et~al.}
\newblock \bibinfo{title}{{The Nuclear Spectroscopic Telescope Array (NuSTAR)
  High-energy X-Ray Mission}}.
\newblock \emph{\bibinfo{journal}{\apj}} \textbf{\bibinfo{volume}{770}}~(2),
  \bibinfo{pages}{103} (\bibinfo{year}{2013}).


\bibitem{Komatsu2009}
\bibinfo{author}{{Komatsu}, E.} \emph{et~al.}
\newblock \bibinfo{title}{{Five-Year Wilkinson Microwave Anisotropy Probe
  Observations: Cosmological Interpretation}}.
\newblock \emph{\bibinfo{journal}{\apjs}} \textbf{\bibinfo{volume}{180}},
  \bibinfo{pages}{330--376} (\bibinfo{year}{2009}).

  .

\bibitem{Ajello2020}
\bibinfo{author}{{Ajello}, M.} \emph{et~al.}
\newblock \bibinfo{title}{{The Fourth Catalog of Active Galactic Nuclei
  Detected by the Fermi Large Area Telescope}}.
\newblock \emph{\bibinfo{journal}{\apj}} \textbf{\bibinfo{volume}{892}}~(2),
  \bibinfo{pages}{105} (\bibinfo{year}{2020}).


\bibitem{Quinn1996}
\bibinfo{author}{{Quinn}, J.} \emph{et~al.}
\newblock \bibinfo{title}{{Detection of Gamma Rays with E > 300 GeV from
  Markarian 501}}.
\newblock \emph{\bibinfo{journal}{\apjl}} \textbf{\bibinfo{volume}{456}},
  \bibinfo{pages}{L83} (\bibinfo{year}{1996}).


\bibitem{Abdo2011}
\bibinfo{author}{{Abdo}, A.~A.} \emph{et~al.}
\newblock \bibinfo{title}{{Insights into the High-energy
  {\ensuremath{\gamma}}-ray Emission of Markarian 501 from Extensive
  Multifrequency Observations in the Fermi Era}}.
\newblock \emph{\bibinfo{journal}{\apj}} \textbf{\bibinfo{volume}{727}}~(2),
  \bibinfo{pages}{129} (\bibinfo{year}{2011}).


\bibitem{Aleksic2015}
\bibinfo{author}{{Aleksi{\'c}}, J.} \emph{et~al.}
\newblock \bibinfo{title}{{Multiwavelength observations of Mrk 501 in 2008}}.
\newblock \emph{\bibinfo{journal}{\aap}} \textbf{\bibinfo{volume}{573}},
  \bibinfo{pages}{A50} (\bibinfo{year}{2015}).


\bibitem{Ahnen2017}
\bibinfo{author}{{Ahnen}, M.~L.} \emph{et~al.}
\newblock \bibinfo{title}{{Multiband variability studies and novel broadband
  SED modeling of Mrk 501 in 2009}}.
\newblock \emph{\bibinfo{journal}{\aap}} \textbf{\bibinfo{volume}{603}},
  \bibinfo{pages}{A31} (\bibinfo{year}{2017}).


\bibitem{Arbet-Engels2021-II}
\bibinfo{author}{{Arbet-Engels}, A.} \emph{et~al.}
\newblock \bibinfo{title}{{Long-term multi-band photometric monitoring of Mrk
  501}}.
\newblock \emph{\bibinfo{journal}{arXiv e-prints}}
  \bibinfo{pages}{arXiv:2109.03205} (\bibinfo{year}{2021}).


\bibitem{Smith2009}
\bibinfo{author}{{Smith}, P.~S.} \emph{et~al.}
\newblock \bibinfo{title}{{Coordinated Fermi/Optical Monitoring of Blazars and
  the Great 2009 September Gamma-ray Flare of 3C 454.3}}.
\newblock \emph{\bibinfo{journal}{arXiv e-prints}}
  \bibinfo{pages}{arXiv:0912.3621} (\bibinfo{year}{2009}).




\end{thebibliography}
\end{document}